\documentstyle[emulateapj,10pt,apjfonts,psfig]{article}
\def\snr{G0.9+0.1}
\def\cxo{{\em Chandra}}
\def\src{CXOU~J174722.8--280915}
\def\etal{{\rm et~al.\ }}

\def\kms{km~s$^{-1}$}

\lefthead{GAENSLER, PIVOVAROFF \& GARMIRE}
\righthead{CHANDRA OBSERVATIONS OF THE PULSAR NEBULA IN 
SNR~\snr}

\begin{document}
\title{{\em Chandra}\ observations of the pulsar wind nebula in
SNR~\snr}
\submitted{Accepted to {\em ApJ (Letters)}\ on 26 Jun 2001}
\author{B. M. Gaensler\altaffilmark{1,2}, M. J. Pivovaroff\altaffilmark{1,3}
and G. P. Garmire\altaffilmark{4}}
\altaffiltext{1}{Center for Space Research, Massachusetts Institute of
Technology, 70 Vassar Street, Cambridge, MA 02139; bmg@space.mit.edu}
\altaffiltext{2}{Hubble Fellow}
\altaffiltext{3}{Current address: Therma-Wave Inc., 1250 Reliance Way,
Fremont, CA 94539}
\altaffiltext{4}{Astronomy and Astrophysics Department, Pennsylvania 
State University, 525 Davey Laboratory, University Park, PA 16802}

\hspace{-5cm}
\begin{abstract}

We present observations with the {\em Chandra X-ray Observatory} of the
pulsar wind nebula (PWN) within the supernova remnant \snr.
At \cxo's high resolution, the PWN has a clear axial
symmetry; a faint X-ray point source
lying along the symmetry axis possibly corresponds to the pulsar
itself. We argue that the nebular morphology can be explained in terms
of a torus of emission in the pulsar's equatorial plane
and a jet directed along the pulsar spin axis, as is seen
in the X-ray nebulae powered by other young pulsars. A bright clump
of emission within the PWN breaks the axisymmetry and
may correspond to an intermediate-latitude feature in the pulsar
wind. 

\end{abstract}


\section{Introduction}
\label{sec_intro}

When a pulsar's relativistic wind is
sufficiently confined, a 
synchrotron-emitting pulsar wind nebula (PWN) is produced
(\cite{gae01b}). Because of the short synchrotron lifetimes
of high-energy electrons, X-ray emission from a PWN
directly traces the current energetics of the pulsar.
The spectral and morphological characteristics of an X-ray PWN
can thus reveal the structure and composition of
the pulsar wind, and possibly also the orientation
of the pulsar's spin axis and/or velocity vector.
For a PWN in which no
pulsar has yet been detected, this nebular emission
is our only insight into the location and energetics
of the central source. 

At the center of the supernova remnant (SNR)~\snr\ is
one of the many PWNe in which no pulsar has yet been seen.
This PWN was first identified by its radio emission
(\cite{hb87}), but has recently
been detected in X-rays using
{\em BeppoSAX}\ (\cite{msi98}; \cite{smib00}).
In these observations, X-ray emission from the PWN
has a power-law spectrum with
photon index $\Gamma = 2.0\pm0.3$
(where $N \propto E^{-\Gamma}$), 
with an unabsorbed flux density $f_x = 6.6^{+1.3}_{-0.8}\times
10^{-12}$~erg~s$^{-1}$~cm$^{-2}$
in the energy range 2--10~keV. The inferred column
density is very high, $N_H = (1.1\pm0.2)\times10^{23}$~cm$^{-2}$, 
accounting for the lack
of any X-ray emission from the surrounding SNR and suggesting
an approximate distance $\sim$10~kpc. The
size of the SNR shell then implies an age
$\tau = 1-7$~kyr (\cite{msi98}). From the X-ray luminosity
of the PWN, Sidoli \etal\ (2000\nocite{smib00})
estimate that the central pulsar has
a spin-down luminosity
$\dot{E} \sim 1.5 \times 10^{37}$~erg~s$^{-1}$,
and a spin-period $P \sim$~80--190~ms.

We here present higher resolution observations of this source
with the {\em Chandra X-ray Observatory}. This experiment
had two aims: to study
the morphology of the PWN and to search for emission from
the central pulsar.

\section{Observations and Analysis}

Observations of \snr\ were made with the Advanced CCD
Imaging Spectrometer (ACIS) aboard \cxo.    ACIS is an array of ten
1024$\times$1024-pixel  CCDs fabricated by MIT Lincoln Laboratory,
and is the result of a collaboration between MIT and PSU.
These X-ray sensitive devices have large detection efficiency
(10--90\%) and moderate energy resolution ($E/\Delta E \sim 10-50$) 
over a 0.2--10.0~keV
passband.  The CCDs have a scale
of $0\farcs492$ per pixel, well-matched to the on-axis point-spread
function (FWHM $\la1''$).  
The instrument and its calibration are described
in detail by Burke \etal\ (1997\nocite{bgb+97}) and Bautz
\etal\ (1998\nocite{bpb+98}).

A single exposure of length 35~ks was made on 2000~Oct~27
in the standard ``Timed
Exposure'' mode, with the aim-point located at the center of the
back-illuminated S3 chip.  Data were first processed by the Chandra
Science Center (ASCDS version number R4CU5UPD11.1), and were
subsequently analyzed with CIAO v2.1. A brief interval of high
background was rejected, resulting in a usable exposure time of
34\,741~seconds.  Counts received from energies below 0.5~keV or above
8.0~keV were dominated by background; data meeting these criteria were
excluded from subsequent analysis.

X-ray emission from
the bright star HD~161507 was detected at
the western edge of the S3 chip. The X-ray position
measured for this source agrees with the optical position given in the
TYCHO Reference Catalogue (\cite{hkb+98}) 
to better than $0\farcs5$ in each coordinate.

\section{Results}
\label{sec_results}

In Figure~\ref{fig_xray_radio}\ we compare a broad-band
(0.5--8.0~keV) image of SNR~\snr\ 
to a radio image formed
from a re-analysis of the Very Large Array (VLA) observations 
described by Helfand \& Becker (1987\nocite{hb87}). Other than
point sources distributed throughout the field, the only emission
seen in the \cxo\ data is an extended X-ray source coincident with
the central radio PWN. The brightest part of the X-ray nebula
is approximately $1' \times 1'$ in extent, and lies within the
eastern half of the radio PWN. Fainter X-ray emission is seen
to the north, south and west, the full extent of the X-ray nebula
corresponding well to that seen in the radio. No X-ray
emission can be seen corresponding to the radio shell of diameter
$7'-8'$ which surrounds the central nebula.

Figure~\ref{fig_xray} shows the brighter parts of the X-ray nebula
at higher resolution. This region shows a definite axisymmetry, with
the reflection axis running at an angle of $\approx 165^\circ$
(measured counterclockwise from north). Four main features
are seen in this region: 

1. Near the center of this region,
a bright elliptical clump of emission, of dimensions $5'' \times 8''$.
The right panel of Figure~\ref{fig_xray} shows a higher-resolution
image of this source, showing a bright region along the
southwestern edge of this clump. 

2. $10''$ to the south of this clump, an unresolved source 
of signal-to-noise $\sim4\sigma$. The position of this
source is (J2000)
RA~$17^{\rm h}47^{\rm m}22\fs8$, Dec~$-28^{\circ}09'15\farcs0$,
with approximate uncertainties of $\pm0\farcs5$ in each coordinate.
In further discussion, we refer to this source as \src.

3. Running east-west across the nebula,
a semi-circular arc of radius $25''-30''$ which is
bisected by the symmetry axis.

4. In the southern half of the nebula,
an elongated jet-like feature of length $35''-40''$ which
lies along the symmetry axis. 

After applying
a background correction,
$2332\pm66$ counts were extracted for the overall nebula, corresponding to
a count-rate of $67.1\pm1.9$~cts~ks$^{-1}$.
Fitting a power-law spectrum to these data implies
an absorbing hydrogen column $N_H = (1.6\pm0.2) \times 10^{23}$~cm$^{-2}$,
a photon index $\Gamma = 2.3\pm0.4$ and an unabsorbed flux density
in the range 2--10~keV of 
$f_x = (9.6^{+1.7}_{-1.5})\times10^{-12}$~erg~s$^{-1}$~cm$^{-2}$
(all uncertainties $\pm1\sigma$).
For this fit we find $\chi^2= 123$ for 138 degrees of freedom.

We lack sufficient counts to perform accurate spectral
fits to sub-regions of the nebula. To look for
spectral variations, we define the hardness
ratio $H$ to be the ratio of incident counts
at energies 5--8~keV to those with energies 3--5~keV (less than 5\%
of counts from the nebula originate at energies below 3~keV).
The corresponding count-rates and values of $H$ for the different 
components are given in Table~\ref{tab_data}.  

The excess
of photons along the southwestern edge of the bright
central clump suggests the possible
presence of an embedded point source. In order to constrain
the flux of such a source,
we generated
a simulated point source at the position of the peak count-rate in the
clump, at coordinates (J2000) RA~$17^{\rm h}47^{\rm m}22\fs9$,
Dec~$-28^{\circ}09'05\farcs6$.  We then multiplied this point source by
a scaling factor, and subtracted the result from the image.  By
steadily increasing the scaling factor until the residual image showed
a significant deficit at the position of the simulated point source, we
can put an upper limit of $\sim$55~counts on the contribution from 
such a source, corresponding to a maximum count-rate of
1.6~cts~ks$^{-1}$. This brightness enhancement is only
seen at high energies; for the 11 photons falling at
the peak position within the clump, 
the hardness ratio is $H=4.5^{+24.3}_{-3.4}$
(where uncertainties are at 90\% confidence, and have been
derived using the prescription of Gehrels 1986\nocite{geh86}). 

\section{Discussion}

The overall morphology of the X-ray nebula matches that of the radio
nebula, making it certain that we have detected X-ray emission from the
PWN associated with \snr.  Our spectral fit to the overall nebula is
comparable to that obtained by Sidoli \etal\ (2000\nocite{smib00}),
although with values of $N_H$, $\Gamma$ and $f_x$ all 1--2$\sigma$
higher than for the {\em BeppoSAX}\ data.
The region of brightest X-ray emission (shown in the right
panel of Fig~\ref{fig_xray}) is offset from the central peak
of the radio PWN by $\sim30''$ to the north-east. This presumably
reflects a change in the nebular conditions on a time-scale intermediate
between the X-ray and radio synchrotron lifetimes. Specifically,
this offset is most
likely due either to motion of the pulsar away from its birth site or
to interaction between the PWN and the SNR reverse shock (\cite{che98}).

The hardness ratios listed 
in Table~\ref{tab_data} provide no evidence for any significant spectral
variations in the PWN; the ``jet'' feature has a higher hardness
ratio than the rest of the PWN,
but only at the $1\sigma$ level. As a point of reference,
we find empirically that for the region of the CCD which we are considering,
a power-law spectrum with absorbing column $N_H = 1.5\times10^{23}$~cm$^{-2}$
and hardness ratio $H$ has a photon index $\Gamma \approx 4.8 -5H$. 
The values of $H$ for all components are therefore consistent with
the mean photon index $\Gamma = 2.3$ inferred from spectral fitting.

We are unable to derive an upper limit on the flux for the shell
component of the SNR: the shell (as delineated by its radio
morphology) fills the entire CCD,  so that a background correction is
not possible.  However, given the high absorbing column
measured towards the PWN, it is unlikely that any thermal emission from
the SNR shell is detectable. This can be demonstrated by considering
the young ($\la$2000~yr) and X-ray bright ($L_x$ [1--10~keV]~$\approx 3
\times 10^{35}$~erg~s$^{-1}$) SNR~G27.4+0.0 (Kes~73), which has an
X-ray spectrum approximated by thermal bremsstrahlung at a temperature
$kT = 0.8$~keV (\cite{gv97}). If we assume this same spectrum for the
shell of \snr, then at a distance of 10~kpc and for a foreground column
$N_H \approx 1.5\times10^{23}$~cm$^{-2}$, the expected count-rate for
the S3 CCD is $\approx31$~cts~ks$^{-1}$ in the energy range 0.5--8~keV.
Thus even if the shell is as bright as that for G27.4+0.0,
we expect it to produce only $\sim$1000 counts spread over the entire
CCD. This would not be detectable when compared to the 20\,000--30\,000
background counts accumulated over the same area.

\subsection{A Central Pulsar?}

X-ray observations of PWNe often reveal the
pulsar itself, in the form of a centrally-located point source.
In the case of \snr, there are two possible locations
for such a source --- the unresolved source \src,
and a possible point source embedded in the clump $10''$ to the north
of this.

Unlike most of the point sources
seen elsewhere in the field, \src\ is only seen at 
energies above 3~keV, indicating
that it is highly absorbed. We can
thus rule out the possibility that \src\ is a foreground
star.  We next consider the possibility that \src\ is
a background AGN. Adopting
a power-law spectrum of photon index $\Gamma = 1.2$ (\cite{mcba00})
and an absorbing column $N_H \approx 2 \times 10^{23}$~cm$^{-2}$ through
the entire Galaxy, the
count-rate measured for \src\
corresponds to an unabsorbed 2--10~keV flux density
of $\sim6\times10^{-14}$~erg~cm$^{-2}$~s$^{-1}$. There
are approximately 20--50 background sources per square degree of this
flux density or greater (\cite{mcba00}; \cite{trn+01}); the probability of one
such source falling within the boundaries of the X-ray PWN
by chance is $\sim(6-14)\times10^{-3}$. 
We thus think it likely that \src\ is associated with the PWN and SNR.

If \src\ is a young central pulsar, we expect X-ray emission
from it to correspond either to
non-thermal emission from the magnetosphere or
to modified blackbody emission from the neutron star surface (e.g.\ 
\cite{bt97}).
We first consider a blackbody spectrum:
a neutron star of age $10^3-10^4$~yr is expected
to have a surface temperature $kT \approx 0.1$~keV (e.g.\ \cite{vle95}), which
for $N_H = 1.5 \times10^{23}$~cm$^{-2}$ corresponds to $H=2\times10^{-8}$,
clearly in disagreement with the observed value $H=0.54^{+1.05}_{-0.37}$.
Invoking an extremely high temperature $kT = 1$~keV gives
$H=0.4$, but requires a very small emitting area
(inferred blackbody radius of 70~m) in order to match the observed count-rate.
Alternatively, if we assume a power-law spectrum 
for the emission from \src, then for
$\Gamma = 1.5$ (typical for magnetospheric emission;
\cite{bt97}) and $N_H = 1.5 \times10^{23}$~cm$^{-2}$, we expect 
a hardness ratio $H = 0.67$, in
good agreement with the value in Table~\ref{tab_data}.
From the limited spectral information available, we therefore conclude that
the X-rays from \src\ are most likely non-thermal in origin.
For a distance $10d_{10}$~kpc, the isotropic
unabsorbed luminosity for the source
is then $\sim6\times10^{32}d_{10}^2$~erg~s$^{-1}$ in the energy range
2--10~keV.

The possibility that the pulsar is embedded in the clump $10''$ to the
north of \src\ must also be considered, especially given the fact that
the brightest part of the clump has an unusually hard spectrum. While
magnetospheric emission from an embedded pulsar ($\Gamma = 1.5$) should
indeed have a harder spectrum than the surrounding nebula ($\Gamma =
2.3$), the expected hardness ratio at this position is
$0.49 < H < 0.61$
for the upper limit on the point-source count rate which we determined in
\S\ref{sec_results}. 
This is significantly lower than the value 
$H=4.5^{+24.3}_{-3.4}$
determined at the brightest point in the clump, suggesting that this
bright region is not an embedded point source but  rather represents
unresolved structure and/or statistical fluctuations within this
region.

In summary, \src\ provides the only clear evidence for
a hard point source within the PWN, and we therefore
consider this source to be the best candidate for emission
from a central pulsar. In this case, we expect
its X-ray emission to be strongly pulsed at the pulsar
spin-period; unfortunately our data lack the time-resolution and
sensitivity to test this hypothesis. The X-ray
luminosity of \src\ is $\sim0.5$\% of the total X-ray luminosity of the
PWN in the energy range 2--10~keV, which is somewhat lower than
the ratio of magnetospheric pulsar emission to surrounding nebular
emission seen in PWNe whose pulsars have been identified (\cite{bt97}).

\subsection{Nebular Morphology}

The X-ray morphology of the PWN is strikingly axisymmetric.
This symmetry axis presumably reflects the orientation of
its associated pulsar, and likely indicates either
the direction of the pulsar's motion or the orientation of the
pulsar spin axis.

\begin{figure*}[hbt]
\vspace{1cm}
\centerline{\psfig{file=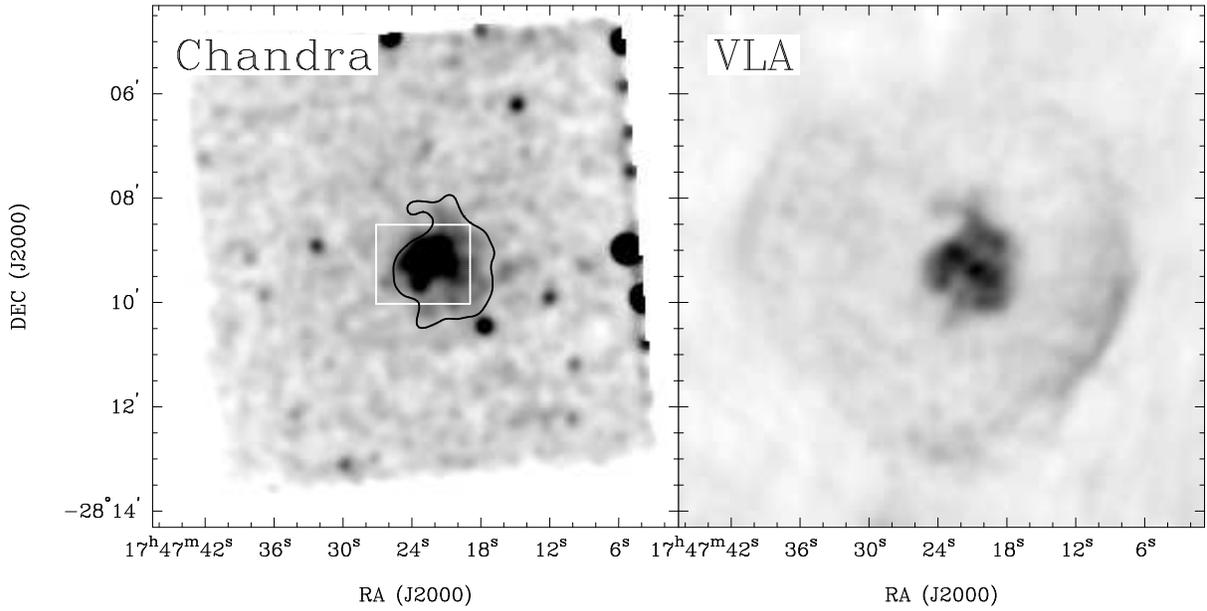,height=8cm,angle=270}}
\caption{X-ray and radio comparison of SNR~\snr. On the left is shown
the \cxo\ image of the entire S3 chip in the energy range 0.5--8.0~keV,
corrected with an exposure map weighted by the spectrum of the
central nebula, and then
convolved with a gaussian of FWHM $15''$. On the right
is shown a 1.5-GHz VLA image of the same region, with a spatial resolution
of $11'' \times 15''$. 
The black contour drawn on the X-ray image
corresponds to 1.5-GHz radio emission at the level of 30~mJy~beam$^{-1}$,
and delineates the outer boundary of the radio PWN;
the white box marked on the X-ray image
shows the extent of the region enclosed by the left panel
of Figure~\ref{fig_xray}.}
\label{fig_xray_radio}
\end{figure*}

In the former case, the pulsar's projected direction of motion is
presumably in the direction $15^\circ$ west of north, and the
semi-circular arc represents a bow-shock produced where the pulsar wind
comes into pressure equilibrium with surrounding material.  However, it
is then difficult to understand the jet-like feature extending to the
south of \src.  The radio/X-ray spectrum of the PWN shows a steepening
at a frequency $\nu_b \sim 10^{11}-10^{12}$~Hz (\cite{smib00}); if this
spectral break results from synchrotron cooling, then electrons
emitting at a frequency $\nu = \nu_b$ must have a radiative lifetime
$t_r \sim \tau = 1-7$~kyr.  Since $t_r \propto \nu^{-1/2}$, electrons
emitting at 5~keV must have a synchrotron lifetime $t_r \sim
(3-10)\times10^{-4}~\tau \sim 0.3-7$~yr.  Even for an extreme velocity
$V\sim2000$~\kms, the pulsar can only travel $<1$\% of the length of
the jet-like feature in this short time. The jet-like feature therefore
cannot correspond to a wake of synchrotron-emitting particles, and 
has no clear explanation
within the bow-shock interpretation.  Furthermore, we
expect the bow-shock morphology to be clearly apparent at radio
wavelengths, as is seen for the PWNe around other high velocity pulsars
(e.g.\ \cite{fggd96}; \cite{ocw+01}). However,  it is clear from
Figure~\ref{fig_xray_radio} that in this case the radio emission is
amorphous and clearly extends well to the north of the X-ray arc. We
therefore think a bow-shock to be an unlikely interpretation of the
data.

The alternative is that the X-ray nebula's symmetry axis corresponds to
the pulsar spin-axis. Such a possibility has been proposed for the
PWNe associated with the Crab and Vela pulsars (\cite{hss+95};
\cite{hgh01}), and possibly also for PSRs~B1509--58 and B0540--69 
(\cite{bb97}; \cite{gw00}).  In all these cases, it has been argued
that the pulsar's X-ray PWN has two main components: a torus
representing the region where an equatorially-concentrated wind
terminates and shocks, and a pair of jets running perpendicular to this
torus which correspond to a collimated wind directed along the
pulsar spin-axis.

At the available sensitivity and resolution, it is reasonable to
propose that the X-ray PWN within SNR~\snr\ has a similar underlying
morphology: the semi-circular arc can be interpreted as half of
an equatorial torus, while the elongated feature running along the 
symmetry axis
would then be a polar jet. In this interpretation the
pulsar is expected to lie at the center of the arc and
along the jet;
the fact that \src\ sits at this position supports 
the argument that this point source is indeed the pulsar.

\begin{figure*}[hbt]
\centerline{\psfig{file=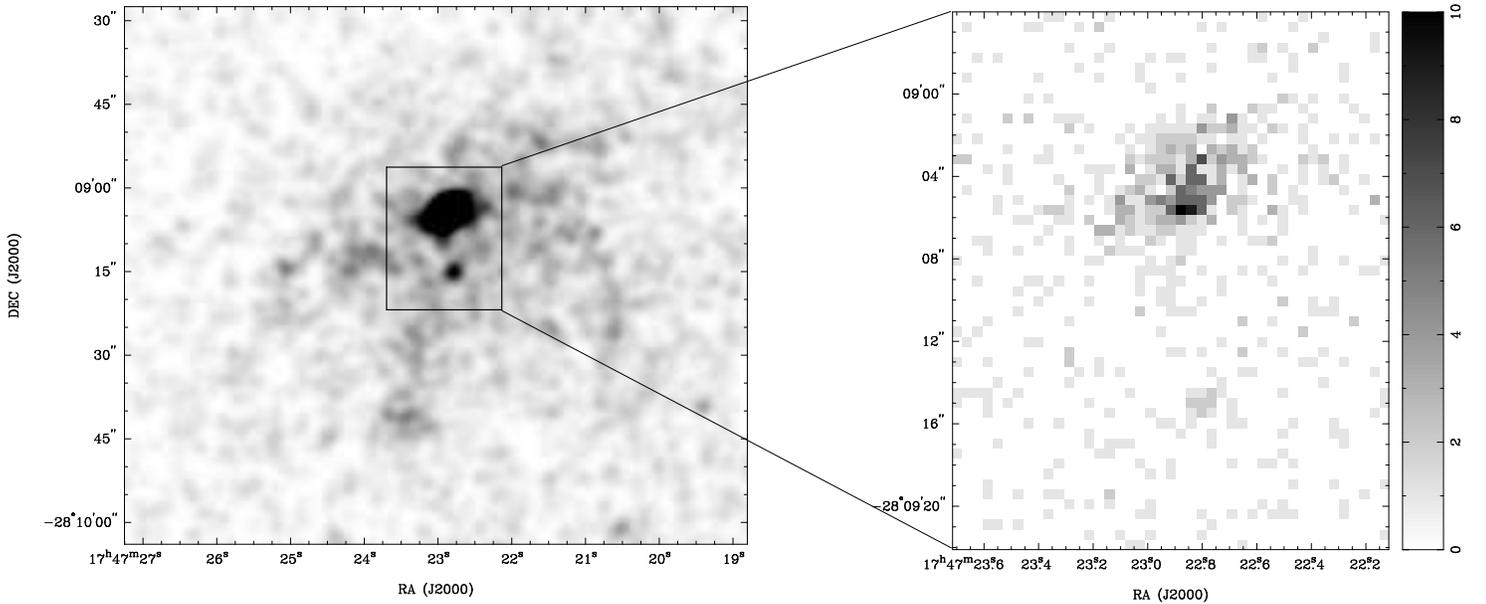,height=8cm}}
\caption{\cxo\ observations of the X-ray PWN within SNR~\snr\ in
the energy range 3.0--8.0~keV. On the
left is shown the inner region of the nebula, exposure-corrected
and then smoothed with
a gaussian of FWHM $3''$. On the right are shown the
central clump and unresolved source (\src) within this inner region.
No exposure correction or smoothing
has been applied to this latter image; the scale shown
at right is in units of raw counts.}
\label{fig_xray}
\end{figure*}

If the torus is circular when viewed along the polar
axis, its ratio of projected minor to major axes of $\sim0.75$
(as measured from the left panel of Fig~\ref{fig_xray})
implies an inclination of the pulsar spin axis to the line-of-sight
of $\theta\sim40^\circ$. The radius of the torus is then $R\sim 1.2d_{10}$~pc
and the length of the jet 
is $L\sim2.5d_{10}$~pc. The absence
of the other half of the torus and of a counter-jet,
as are seen also for the PWN around the Vela pulsar (\cite{hgh01}),
can be accounted for by Doppler boosting: for wind speeds $>c/3$
(\cite{kc84a}) and $\Gamma = 2.3$, the approaching components of
the jet (torus) will be boosted
to a brightness a factor $r>6$ ($r>4$) over
receding material. The observed ratios
of surface brightness of the jet and arc compared
to any counter-jet and counter-arc, respectively, are
both $r \ga 5$, consistent with this interpretation.
The light-travel time along the jet is $\sim8d_{10}$~yr.
Provided that $d_{10} \la 1$ and that the outflow velocity is close to $c$,
this is consistent with the upper end of the range of
synchrotron-emitting lifetimes inferred above.

The scale of features seen in this PWN
is considerably larger than for the nebulae
around the Crab ($R \sim 0.4$~pc, $L \sim 0.6$~pc; \cite{hss+95})
or Vela ($R \sim 0.03$~pc, $L \sim 0.01$~pc; 
\cite{hgh01})\footnote{The jet we refer to in the case of the Vela pulsar
is the $10''$-long feature aligned with the pulsar spin axis. A
much longer ``jet'' has been reported by Markwardt
\& \"{O}gelman (1995\nocite{mo95}),
but does not form part of the axisymmetric morphology
of the inner X-ray nebula.} pulsars, but are comparable
to that of the X-ray PWN surrounding PSR~B1509--58,
($R\sim1.5$~pc, $L > 7$~pc; \cite{bb97}; \cite{tkyb96}).
In this latter case, the large extent of the PWN
is thought to be due to the
low density environment ($n \la 0.01$~cm$^{-3}$;
\cite{bha90};  \cite{gbm+98}). A similarly low
density for \snr\ is consistent with its faint radio shell.

The bright clump seen to the north of \src\ cannot be explained within
the context of a ``torus + jets'' geometry.  The clump is clearly 
a discrete structure superimposed on the fainter nebular
background, and is not a geometric effect resulting from overlapping
emission from torus and jet components.  The fact that the clump is
offset slightly to the east of the symmetry axis,
and is elongated along a different axis to that of the
PWN, suggests that it is not part of the overall axisymmetric
structure; it may represent an intermediate-latitude feature where the
pulsar wind is shocked or compressed. Such compact features 
may be highly time-variable, as has been seen in optical
and X-ray observations of the Crab and Vela PWNe respectively
(\cite{hes01}; \cite{pksg01}).

Regardless of any detailed interpretation for the nebular
morphology seen here,
the data demonstrate that the flow
of energy away from the central pulsar in SNR~\snr\ is distinctly anisotropic,
and add support to the hypothesis that axial/equatorial
wind morphologies are generic amongst the pulsar population.
It is clear that the high spatial resolution now
available in X-rays with \cxo\
can play an important role in determining
how young pulsars deposit their energy into their environments.

\begin{acknowledgements}

We are grateful to Mark Bautz and the ACIS Team for inviting us to work
on this project, and thank Fred Baganoff, Dave Pooley and Pat Slane for
useful discussions.  B.M.G. acknowledges the support of NASA through
Hubble Fellowship grant HST-HF-01107.01-A awarded by the Space
Telescope Science Institute. M.J.P. acknowledges the support of NASA
contracts NAS8-37716 and NAS8-38252.

\end{acknowledgements}

\bibliographystyle{apj1}
\bibliography{journals,modrefs,psrrefs,crossrefs}


\begin{deluxetable}{lcc}
\tablewidth{500pt}
\tablecaption{Count rates and hardness ratios for components
of the  X-ray nebula.\label{tab_data}}
\tablehead{
\colhead {Region} & \colhead{Count Rate (cts~ks$^{-1}$)\tablenotemark{a}} & 
\colhead{Hardness ratio\tablenotemark{b}} } 
\startdata
Entire nebula & $67.1\pm1.9$ & $0.51\pm0.05$ \nl
``Arc'' & $21.0\pm1.0$ & $0.54^{+0.09}_{-0.08}$ \nl
``Clump'' & $10.0\pm0.6$ & 0.49$^{+0.12}_{-0.10}$ \nl
``Jet'' & $3.8\pm0.4$ & $0.68^{+0.28}_{-0.20}$ \nl
\src & $0.37^{+0.14}_{-0.11}$ & $0.54^{+1.05}_{-0.37}$ \nl
\enddata
\vspace{-2mm}
\tablenotetext{a}{Count rates are for the energy range 3--8~keV,
and have been corrected for background; uncertainties
are $\pm1\sigma$.}
\tablenotetext{b}{Defined as the
ratio of counts with energies 5--8~keV to those
with energies 3--5~keV. Uncertainties are all at 90\% confidence,
and have been
determined using the formulae given by Gehrels (1986\nocite{geh86}).}
\end{deluxetable}

\clearpage

\end{document}